\documentclass[anonymous=false, draft=false]{iscram}
\usepackage{inputenc}
\usepackage{tikz}
\usepackage{fancyvrb}
\usepackage{listings}
\usepackage{enumitem}
\usepackage{tcolorbox}
\usepackage{multicol}
\usepackage{siunitx}
\usepackage{graphics}
\usepackage{multirow}
\usepackage{booktabs}
\usepackage{graphicx}
\usepackage{float}
\usepackage{caption}
\usepackage{subcaption}
\graphicspath{{./figures/}}
\newcommand{\rurl}[1]{\href{http://#1}{#1}}
\iscramset{
  title={
    Integrating Social Media into a Pan-European
    Flood Awareness System: A Multilingual Approach
  },
  short title={
    Integrating Social Media into a Pan-European Flood Awareness System
  },
  author={
    short name={Lorini},
    full name=Valerio Lorini,
    affiliation={
      European Commission, Joint Research Centre (JRC), Ispra, Italy \\
      Universitat Pompeu Fabra, Barcelona, Spain \\
      {valerio.lorini@ec.europa.eu}
    },
  },
  author={
    full name=Carlos Castillo,
    affiliation={
      Universitat Pompeu Fabra, Barcelona, Spain \\
      chato@acm.org
    },
  },
  author={
    short name={Francesco Dottori},
    full name=Francesco Dottori,
    affiliation={
      European Commission, Joint Research Centre (JRC), Ispra, Italy \\
      francesco.dottori@ec.europa.eu
    },
  },
  author={
    short name={Milan Kalas},
    full name=Milan Kalas,
    affiliation={
      KAJO, Bytca, Slovakia \\
      milan.kalas@kajoservices.com
    },
  },
  author={
    short name={Domenico Nappo},
    full name=Domenico Nappo,
    affiliation={
      European Commission, Joint Research Centre (JRC), Ispra, Italy \\
      domenico.nappo@gmail.com
    },
  },
  author={
    short name={Peter Salamon},
    full name=Peter Salamon,
    affiliation={
      European Commission, Joint Research Centre (JRC), Ispra, Italy \\
      peter.salamon@ec.europa.eu
    },
  },
}

\begin{document}

\maketitle
\makeatletter
\makeatother

\abstract{
  This paper describes a prototype system that integrates social media analysis into the European Flood Awareness System (EFAS).
  This integration allows the collection of social media data to be automatically triggered by flood risk warnings determined by a hydro-meteorological model.
  Then, we adopt a multi-lingual approach to find flood-related messages by employing two state-of-the-art methodologies: language-agnostic word embeddings and language-aligned word embeddings.
  Both approaches can be used to bootstrap a classifier of social media messages for a new language with little or no labeled data.
  Finally, we describe a method for selecting relevant and representative messages and displaying them back in the interface of EFAS.
}



%

\section{Introduction}
\label{sec:introduction}

Over millennia humans have developed villages and cities near water bodies, mainly for two reasons:
(i) improvements in agricultural yields due to yearly floods in fertile floodplains leaving nutrient-rich silt deposits behind; and
(ii) people's desire to live near coastlines and river valleys, often on wetlands and backfilling otherwise natural flood buffers.
Due to the closeness of human settlements to rivers and coasts, floods are the natural disasters with the greatest damage potential and the ones that affect the greatest number of people \parencite{ocha2017humanitarian}.

According to recent studies, Global warming is linked to a substantial increase in flood risk over most countries in Central and Western Europe, even if the temperature increases remain close to the lower estimates.
For instance, flood peaks with magnitude as it happens once in 100 years, called 'return period', are projected to double in frequency within the next three decades \parencite{alfieri_global_2015}.\footnote{The return period of an event is the time span it would take to observe one such event on expectation.}

In Europe, the Emergency Response Coordination Centre (ERCC), operating within the European Commission's Civil Protection and Humanitarian Aid Operations department, was set up to support a fast and coordinated response to disasters both inside and outside Europe using resources from the countries participating in the EU Civil Protection Mechanism.
This centre monitors hazards and risks, collects and analyzes real-time information on disasters, prepares plans for the deployment of experts, teams and equipment, and in general coordinates the EU's disaster response efforts. %

The European Flood Awareness System (EFAS) provides real-time information and forecasts about floods to the ERCC as well as to a series of partners including national and regional hydrological services.
EFAS is part of the Copernicus Emergency Management Service (Copernicus EMS), and holds regularly updated flood-related information such as probabilistic medium-range flood forecasts (including short-range flash floods), seasonal forecasts, and impact assessments and early warnings.

Current flood hazard mapping methodologies, such as the one implemented in EFAS, have high scale spatial resolution capacity: $1 \times 1$ kilometer resolution globally and $100 \times 100$ meter resolution in Europe \parencite{dottori_development_2016}.
This combined with state-of-the-art forecasting models provide detailed information about the risk associated with a flood in terms of likelihood, magnitude, timing, and impact \parencite{dottori_operational_2017}.

Due to nature of floods phenomena and its dynamics, it is important that satellites can monitor the flood extent during the flood peak, therefore early warnings are important to allow sufficient lead time for requesting satellite mappings of the area at risk \parencite{dottori2017satellites}.

Over the past decade, social media has emerged as a relevant information source during disasters, prompting researchers from diverse areas to converge on this domain \parencite{castillo_big_2016,palen2016crisis}.
Social media analysis has demonstrated the potential to provide timely, precious information about the spatial (\cite{brouwer_probabilistic_2017,restrepo-estrada_geo-social_2018}) and
temporal \parencite{aiello_integrating_2014} development of a crisis,
as well as supporting the identification of key disaster-related events \parencite{litmus2015}.
Although the value of social media analysis in providing timely data and methods for the analysis of natural hazards has been recognized in previous work, comparatively much less attention has been given to how to integrate social media in a seamless, reliable way with tools for disaster forecasting and monitoring.

Since the geographical domain of EFAS products covers an area where population speaks more than 27 languages, we focused on a multi-lingual system. We therefore use representation of words as vectors in order to exploit probabilistic functions to infer similarities between words, know as word embeddings\parencite{pennington2014glove}.

Our research fills this gap by describing the integration of social media monitoring into a flood monitoring and forecasting platform, enriching hydro-meteorological information with reports from the public.

\begin{figure}[htb]
\centering
\includegraphics[width=\textwidth]{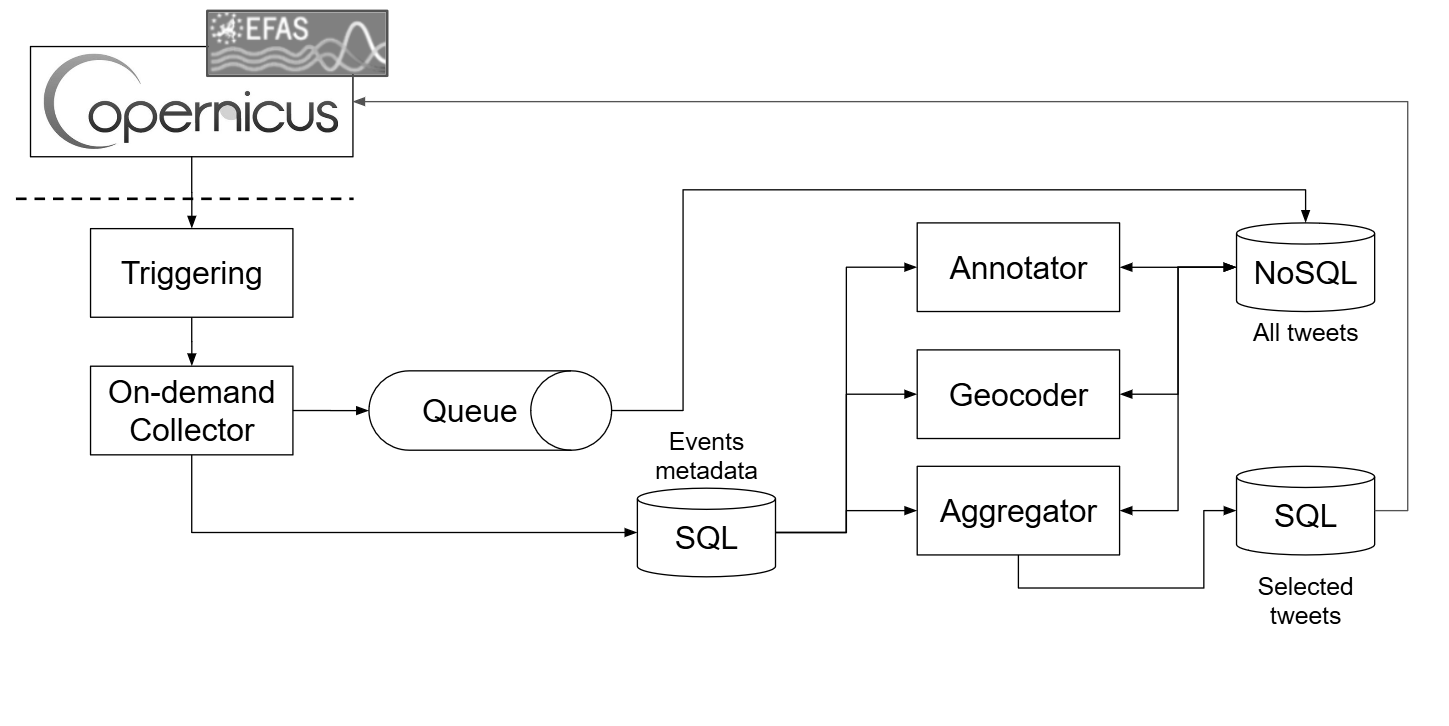}
\caption{Schema of SMFR components}
\label{smfr-schema}
\end{figure}

We developed software for EFAS named Social Media for Flood Risk (SMFR) which provides near-real-time information collected from social media about flood risks and impacts, including examples of messages in social media about it. Figure~\ref{smfr-schema} represents the conceptual schema of SMFR’s components, part of which is described in this paper, and their integration.

Our main contributions are:
\begin{itemize}
\item We integrate social media data collection into the European Flood Awareness System (EFAS) based on EFAS forecasts.
Whenever EFAS rapid risk assessment identifies heightened risk of floods in a certain area, we trigger data collection from social media (in our case, Twitter) respecting API limitations while dealing with the possibility of various events happening at the same time.
A similar mechanism is already in use for triggering pre-tasking of satellite image acquisition in Copernicus EMS
\item We describe a methodology that requires a minimal amount of manual intervention for each additional language, and demonstrate it with four languages being used to bootstrap a classifier for a fifth language.
This methodology is based on Convolutional Neural Networks and its multi-lingual capabilities stem from using either language-agnostic word embeddings, where vectors representing sentences are not dependent on a single language,  or multi-lingual word embeddings \parencite{conneau_word_2017}.
\item We describe an aggregation and selection module that can select representative messages for an area in which flood risk has been predicted.
\item We integrate SMFR into EFAS and demonstrate it during the recent floods affecting Calabria, Italy, in early October 2018. Note that we decided to present only one case for the sake of clarity and conciseness, however SMFR has been tested with additional real cases.
\end{itemize}

The remainder of this paper is organized as follows:
the next section presents related work;
then, three technical sections describe the methods for on-demand data collection, multi-lingual classification, and aggregation and selection.
Finally, we present experimental results, including a case study, followed by our conclusions and future work.

\section{Related Work}
\label{sec:related-work}

\subsection{Background: European Flood Awareness System (EFAS)}
EFAS is part of the Copernicus Emergency Management Systems, and serves the ERCC and EU member states' Civil Protection agencies with forecasts of flood risks. It covers the European Union as  well as several neighbouring areas that are relevant from the perspective of flood risk and EU policies (e.g. all countries of the Danube river basin).
EFAS provides information based on weather forecasts\footnote{EFAS inputs are based on the Ensemble Prediction System from the European Center for Medium Range Weather Forecast (ECMWF) which consists of 51 ensemble member.} and hydrological ensemble predictions obtained by an hydrological model simulations, therefore subjected to uncertainty. For being exploited at its maximum potential it should offer additional supporting information that could be used for prioritizing resources and interventions. EFAS already includes products that go in that direction. For instance pre-tasking satellite mappings is triggered in advance integrating socio-economic data (population, infrastructures, economic losses fused as impact indicators) with flood magnitude and probability (likelihood)\parencite{dottori2017satellites}.

EFAS forecasts have good accuracy level which is constantly monitored\footnote{publicly available at https://www.efas.eu/validation-and-skill-scores}, but Crisis are only declared, formalised and managed by the local authorities and it's up to them to decide what to do with the data provided. The new layer presented helps in the near real time confirmation of floods for areas where a high flood risk was forecasted.
The SMFR layer includes a set of representative tweets and a dot distribution map of social media messages related to floods.

\subsection{Combining authoritative and non-authoritative data}
A recent trend in research on social media on disaster has been to study methodologies for combining non-authoritative and authoritative data in risk assessment.
The non-authoritative data are reports generated by the public, typically posts in social media platforms.
The authoritative data comes from various sensors including meteorological and hydrological ones as well as physical models for creating forecasts with this data.
In previous work, these data have been combined in various ways.

\textcite{litmus2015} describe the LITMUS platform, which collects and filters messages about landslides from various social media platforms and geo-locates them to merge reported events with data from physical sensors referring to the same location.

\textcite{brouwer_probabilistic_2017} use social media combined with authoritative data to define probabilistic and deterministic flood-extent maps.
Information gathered from the public is used to make inferences about rainfall, contributing to interpolate coarse-grained data at a the same resolution as a Local Drainage Direction (LDD) map, in order to determine water levels for different cells.

\textcite{restrepo-estrada_geo-social_2018} use a transformation function for creating a proxy variable for rainfall by analyzing keywords-filtered geo-located social media messages and rainfall measurements from authoritative sources.
The proxy variable is incorporated in a hydrological model for stream-flow estimation.

Our work differs from the previous ones in important ways. First, we do not filter social media posts by flood-related keywords, but rather according to the location of a possible flood based on forecasts from the EFAS system.
Second, we aggregate data using a geocode standard for referencing the subdivisions of countries for statistical purposes, known as Nomenclature of Territorial Units for Statistics\parencite{nuts_2016}. Its granularity is identified by levels, the higher the level, the higher the granularity. In our work we used level 2 (NUTS-2) as the main subdivision.
Third, we cover a large area by automatically processing content in several languages.
The goal of our system is to confirm and bring more detail to the outcomes of an hydrological model.

\subsection{Multi-lingual classification of social media postings}

SMFR is designed to work across multiple languages. The main processing that we do to messages is to determine whether they are relevant to flood risks/impacts or not. This is done through supervised classification, which requires labeled data. However, to work across multiple languages in practice requires to be able to classify messages in languages for which we may not have labeled data yet.

Past research addressed cross-lingual bootstrapping of classifiers for natural disasters detection on twitter \parencite{middleton_social-media_2013} relying on auto{}matic translation to use available models.
Previous works, such as \textcite{li_comparison_2018} has shown how an approach based on \emph{word embeddings} works better than a simpler method based on bag-of-words when generalization is critical, including our case which requires generalizing across languages.
\textcite{vrandecic_cross-lingual_2018} built a statistical-semantic classification model with semantics extracted from BabelNet and DBpedia and compared relevancy classifiers with datasets translated into a single language, as well as with cross-lingual datasets. They have shown how adding semantics increases cross-lingual classification accuracy.

Our work, in contrast, does not require semantic resources, we only leverage on word embeddings for multi-lingual modelling as demonstrated by \textcite{luong_bilingual_2015}.
Indeed, we present and demonstrate two different ways in which word embeddings can be used to perform multi-lingual classification: using language-agnostic word embeddings learnt from a multi-lingual corpus \parencite{pennington2014glove}, and using multi-lingual word embeddings that are aligned across languages \parencite{conneau_word_2017}.
In both cases, we can use labeled data in a set of known languages to bootstrap a classifier for a new language for which no labels are available.

\section{On-Demand Data Collection}
\label{sec:collection}

The data collection should ideally achieve high recall, capturing a large fraction of the relevant information, while at the same time having high precision, avoiding irrelevant information. Both goals usually enter into conflict and trade-offs are necessary.
Previous research has described extensively how data collection from Twitter is done (see, e.g., \cite{imran2015processing}), in this section we focus on the specific aspects of our system, which performs \emph{on-demand data collection}.

\begin{figure}[H]
\centering
\includegraphics[scale=0.45]{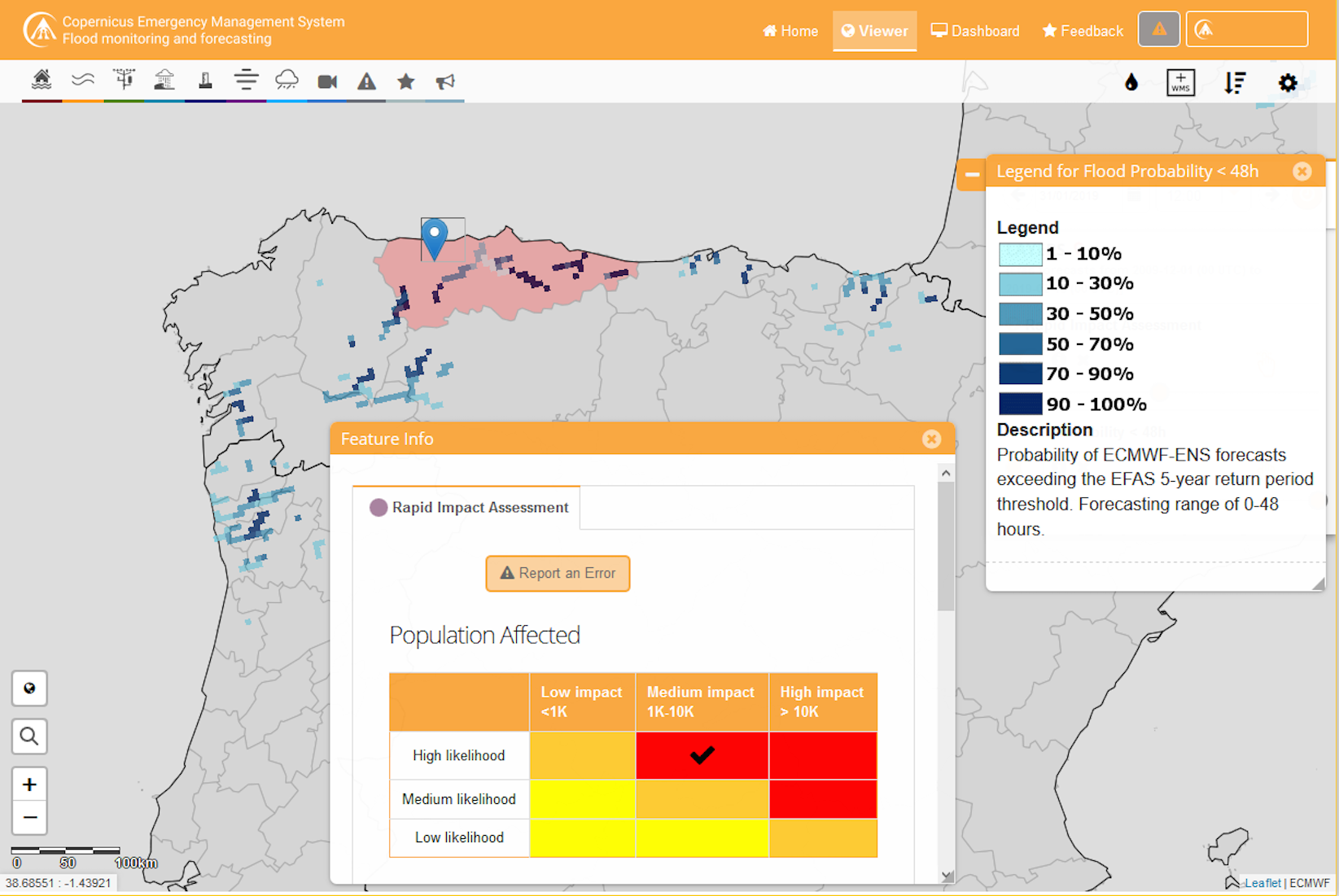}
\caption{Screenshot of EFAS web interface with the layers identifying areas where there is high probability of floods in the following 48 hours and rapid impact assessment}
\label{efas-triggering}
\end{figure}

The key element of our data collection is its triggering mechanism which is done dynamically according to flood forecasts.
EFAS runs two simulations per day, identifying NUTS-2 areas (typically regions or provinces) where there is a high probability of floods impacts in the following 48 hours as shown in figure~\ref{efas-triggering}.
Once the list of NUTS-2 areas is received, the system extracts their coordinates and the names of all cities in the area that have more than a certain number of inhabitants. The coordinates are used for filtering the Tweets by location while the names of cities, in english and local language, translate into a series of 'OR' filters by keywords.
In our current configuration, we set this threshold empirically to 80,000 inhabitants, which tends to capture a handful of cities for each event. A lower threshold is possible as long as it does not generate a large number of city names that exceeds Twitter's API limitations.

Given the limitations in Twitter's API, we use a single connection against Twitter's public streamer at any given time.
Additionally, the public streamer limits queries to up to 400 keywords (each of less than 60 bytes) and up to 25 location rectangles.\footnote{\url{https://developer.twitter.com/en/docs/tweets/filter-realtime/overview} accessed November 2018.}
A \emph{query builder} component groups several active events in order to update a single request respecting these limitations, while keeping information about the single possible event identified by EFAS.
Each collection is kept active for two days after the expected peak time. If the peak time estimation is updated by a new EFAS simulation, the collection's expiration time is extended.
Given that we have a single query, SMFR has to separate the incoming stream into different events according to locations and keywords. If a message belongs to overlapping regions or contains names of cities in different events, the message is copied to all the matching events.

\begin{figure}[H]
\centering
\includegraphics[scale=0.45]{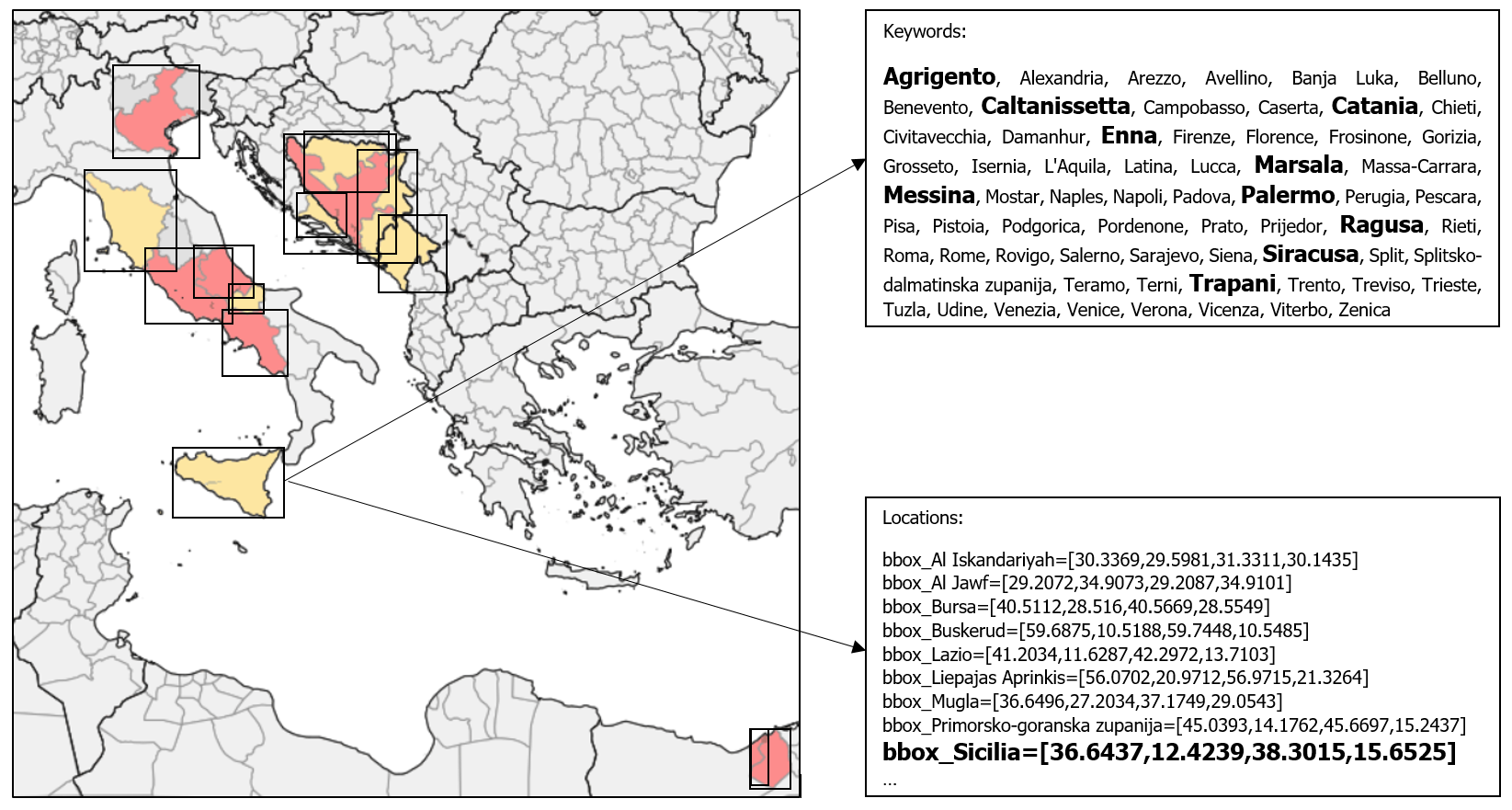}
\caption{Example depicting areas in which EFAS forecasts high risk, which appear in yellow and red color. Each area defines a set of keywords, which are names of cities, and locations, which are bounding boxes (rectangles). These keywords and locations are used to gather information from Twitter.}
\label{on-demand-figure}
\end{figure}

Figure~\ref{on-demand-figure} depicts how rapid risk assessment leads to the definition of a series of keywords (city names) and locations (rectangles containing NUTS-2 areas) for filtering information from Twitter's public streamer.
The areas in yellow and red in the figure are identified as having high risk of flood by EFAS.
Then, the system determines names of cities in each area and bounding boxes surrounding each area.
We have heuristics that merge neighboring areas and de-prioritize smaller cities if their quantity exceeds Twitter's limitations, but in practice these are rarely triggered.

\section{Multi-Lingual Classification}
\label{sec:classification}

The setting used was supervised binary classification. The positive class comprised all messages indicating that ``a flood has just happened or is about to happen'' while the negative class included all other messages.

Training data was labeled using crowdsourcing platform GetHybrid (\url{https://gethybrid.io/}).
The amount of labeled data comprises over 7,000 annotated messages, each one annotated by three annotators independently.
Four training datasets were created, containing labeled tweets in German, English, Spanish, and French.
Each language includes between 1,200 and 2,300 annotated messages, as shown in Table~\ref{table-experiments} (column ``TL'' for Total Labeled) in the section describing the experiments.
This amount of data is typical in automatic classification tasks \parencite{imran2015processing}.

\subsection{Convolutional Neural Networks}
We tested a number of learning algorithms, including Support Vector Machines \parencite{joachims1999svmlight}, which have been shown to be effective for a number of general text classification tasks, and Random Forests, which have been shown to be effective specifically for classifying crisis-related Twitter messages \parencite{imran2014aidr}. In both cases, we represented messages as bag of word unigrams and bigrams, and the performance was better than simpler methods such as  a na\"ive Bayes classifier or a decision tree.

A comparatively newer approach for text classification that has proven to be quite effective is the use of Convolutional Neural Networks (CNN). These have been employed for a number of tasks including sentiment analysis \parencite{severyn2015twitter}.

There are four main operations in every Convolutional Neural Network:(i) Convolution; (ii) Non Linearity, Rectified Linear Unit (ReLU); (iii) Pooling or Sub Sampling;(iv) Classification (Fully Connected Layer). The primary purpose of (i) Convolution is to extract features from the input using a filter smaller in size than the original input. The second (ii) operation ReLU replaces all negative values in the feature map by zero. The (iii) Pooling operation , max pooling in our case, reduces dimensionality by taking the largest element from the rectified feature map within a neighborhood. The (iv) Classification operations, after a series of iteration of the (ii) and (iii) then uses the higher-level features identified to deter a class using several fully connected perceptrons layers.

A Support Vector Machine (SVM) is a classifier defined by a separation hyperplane. Input of the algorithm is the labeled training data, while its output is an optimal hyperplane which places new data into distinct classes.

Random forest classifiers infer a series of decision trees from randomly selected subset of the labeled data. The results are then passed through different decision trees to finally classify the test data.

The specific architecture we used is described in the next section; using it, we performed various experiments using 10-20 epochs of training (passes over the entire training set in randomised order). %

The results were similar in performance to the SVM and Random Forests. However, manual error analysis showed that qualitatively these errors were different.
While in the case of SVM and Random Forests a misclassification, such as a false positive, is usually the result of a word marking flood relevance (e.g., "flood") used in a completely different context (e.g., "my timeline is flooded with messages"), in the case of the neural network we used a misclassification was often semantically related to floods, such as a message referring to the effects of other natural disasters.
The lesson learned from these experiments was that neural networks are better in this problem at capturing semantic characteristics that are relevant for our task of distinguishing flood-related messages.

\subsection{Neural Network Architecture}

Current implementations of convolutional neural networks for text processing tasks tend to have a similar architecture.
They consist of an input layer, a word embedding layer, a series of convolutional and max pooling layers, a dense layer, and an output layer \parencite{liao2016}.

The input layer holds a padded sequence of words with a maximum length $S=100$ words, which is more than sufficient for tweets considering their maximum length is 280 characters.

The word embedding layer converts every word into a low-dimensional vector, typically in the order of a few hundred dimensions (e.g., $D=200$ or $D=300$).
We used two sources of pre-trained word embeddings, as described in the next section.
For each pre-trained word embedding, we considered two configurations-ups: one in which the parameters of the pre-trained word embedding were fixed, i.e., not modifiable while training the neural network, and one in which they were part of the optimization process, i.e., modifiable while training.
In our experiments, best results were obtained when these parameters were fixed, probably because the amount of flood-specific data that we are using for training is small in comparison with the corpora used to create these word embeddings. The results we report on this paper use fixed word embeddings.

The convolution layers collect several word embeddings representing adjacent words and ``summarize'' them into a single vector. The main parameter for the convolutions is the width $C$, which is how many adjacent words to take into account. In the text "flood warning due to heavy rain" using $C=5$, there are two possible convolutions: ``flood warning due to heavy'' and ``warning due to heavy rain.''
The parameter $C$ is determined considering what is the effect of the context on the meaning of a word, and $C=5$ is a typical value. We did not observe any increase in performance with a larger value of $C$, while a smaller value of $C$ may lose contextual information.

The max pooling layer collect a series of $m$ disjoint convolutions as input, and generate a vector of dimension $d < D$ as output. The max pooling step operates differently from the convolution layers in the sense that the windows it uses are disjoint, i.e., non-overlapping.
The purpose of this layer is to reduce the dimensionality of the network for computational purposes and to reduce the chances of overfitting.
In our case we used $m=5$ and $d=128$, which are typical parameters used in text classification.

The final layer is a densely-connected (complete) layer. All the neurons in the last max pooling layer are connected to all the neurons in this dense layer, and all the neurons in this dense layer are connected to the two output neurons.
One of the two output neurons should activate when the example is positive (i.e., a message indicating that a flood has just happened or is about to happen), and the other output neuron should activate when the example is negative (i.e., the message does not indicate that a flood has just happened or is about to happen).

\subsection{Word Embeddings for Multi-Lingual Classification}

The usage of word embeddings allows to incorporate multi-lingual capabilities in two ways: by using language-agnostic word embeddings and by using language-aligned word embeddings.

Our source of language-agnostic word embeddings is GloVe \parencite{pennington2014glove}, which are vectors of dimensionality 200 obtained from a large corpus of tweets containing $27 \times 10^9$ tokens (1.2 million of them unique). While these word embeddings were not developed for multi-lingual tasks, they do incorporate any word present on a tweet in a language-agnostic manner.

Our source of language-aligned word embeddings is MUSE \parencite{conneau_word_2017}, which are vectors of dimensionality 300 obtained from various snapshots of Wikipedia in various languages. For each language, vectors for the 200,000 most frequent tokens are provided, and these vectors have been \emph{aligned} across languages using parallel lists of tens of thousands of words. In the resulting embeddings, two words with the same meaning in different languages are mapped to similar vectors.

In our experiments, presented on the next section, both pre-trained sets of vectors allow to transfer an automatic classifier learnt with labeled data from one language (or a set of languages) into another language with no new labeled data (``cold start'') or with a limited amount of labeled data (``warm start'').

\section{Aggregation and selection}
\label{sec:aggregation}

In our integration with EFAS, flood risk is established by an hydro-meteorological model and our task is to add complementary information that brings a better understanding of the situation on the ground.
Relevant messages are mapped to NUTS-2 areas (Nomenclature of Territorial Units for Statistics, Level 2) either by using explicit coordinates which are rarely present in tweets, or more often, via a text-based geocoder.
Geocoding deals with messages that do not include explicit geographical coordinates, but mention a place name such as landmark or city.
Geocoding uses a Named Entity Recognition tagger to obtain possible locations in a text considering the syntax of the message. It uses a gazetteer in a large database of place names with their corresponding geographical coordinates, and finally it uses a neural networks to infer the correct country and correct gazetteer entry for those places.
We used a library named Mordecai \parencite{halterman2017mordecai} which extracts place names from a piece of text, resolve them to the correct place, and return their coordinates and structured geographic information.

Messages are aggregated at the level of an event, but also at the level of each NUTS-2 area. In both cases, one key operation is to select a representative sub-set of messages.
We do this operation by using the following efficient heuristic based on de-duplication and text centrality:

\begin{enumerate}
\item Select up to 5,000 tweets having at least a 90\% probability of being flood-related; if there are more than 5,000 tweets, select the ones with the highest probability
\item Compute similarities between these tweets
\begin{enumerate}
  \item Consider only pairs having probabilities of being flood-related that differ at most by $0.0001$, exploiting the fact that near-duplicate tweets will be given the same probability by the neural network
  \item Compute edit distance and use it to compute normalized similarity: $1 - \ell(m_1, m_2)/(|m_1| + |m_2|)$ where $\ell(m_1, m_2)$ is the edit distance between the two messages and $|m_1|+|m_2|$ the sum of their lengths.
  \item If the normalized similarity is greater than 0.8, use the timestamps of the tweets to mark the newer tweet as a duplicate of the older tweet
\end{enumerate}
\item Sort all unique tweets by their \emph{multiplicity}, i.e., by the number of duplicates they have, and keep the top 100
\item For these tweets, compute all pair-wise similarities using the same formula as above, and add the similarities for every tweet; this is the \emph{centrality} of the tweet.
\end{enumerate}

Step 1 of the heuristic has the goal of removing messages that are irrelevant.
Step 2 removes near-duplicates which are redundant, but keeps track of how many near-duplicates a tweet has, for the purposes of using redundancy as a signal of importance.
Step 3 applies a well-known lexical centrality heuristic \parencite{erkan2004lexrank}, in which a salient message is one that has content in common with many other messages.

\begin{figure}[htb]
\centering
\includegraphics[scale=0.45]{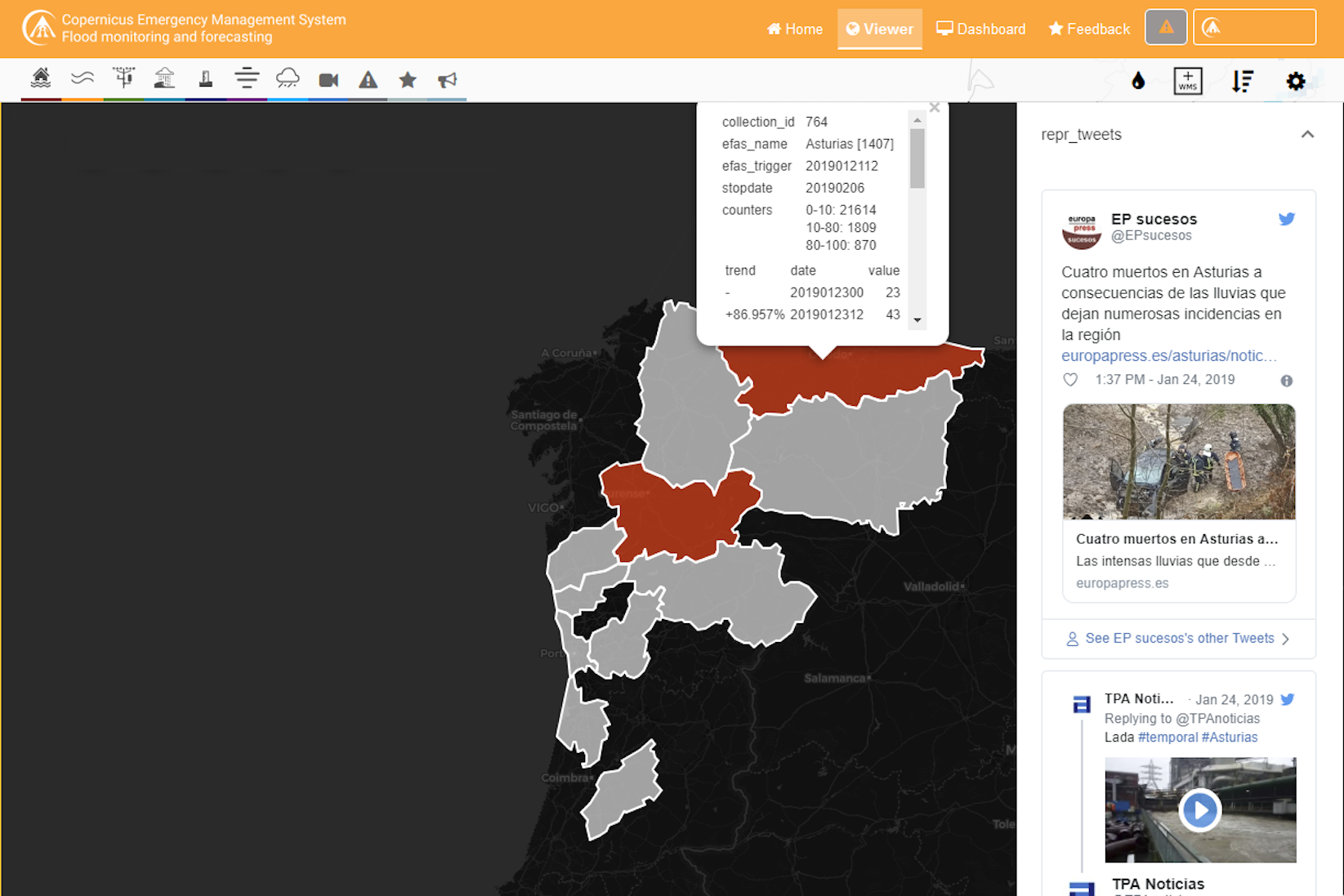}
\caption{Screenshot of EFAS web interface with the layer identifying areas where there is high tweet activity and their most representative tweets on the right side. The basemap has been darkened for better visualization}
\label{smfr-layer}
\end{figure}

Figure~\ref{smfr-layer} shows a first deployment of SMFR on EFAS web interface, it describes how the areas and their most relevant tweets are presented to EFAS users. The NUTS-2 area can be grey (low activity), orange (medium activity) or red (high activity) according to the ratio between numbers of annotated tweets. The triggering that lead to the creation of the collection depicted can be seen in figure~\ref{efas-triggering}

\section{Experimental Results}
\label{sec:experiments}

We describe two types of experimental results.
First, we perform experiments to test the performance of the multilingual-classifier, comparing it with a mono-lingual classifier.
Second, we show a real example of an actual flood event, describing the performance of the on-demand collector, of a multi-lingual classifier that does not use labeled data from the target language, and of the aggregation and selection method.

\subsection{Multi-Lingual Classifier}
\label{subsec:experiments-classifier}

The multi-lingual classifier provides a solution for bootstrapping classifiers in new languages based on labeled data for other languages.
The aim of the experiments is to compare the classifiers trained with and without labeled data for a specific target language.
Performing this evaluation, nevertheless, requires having labeled data to measure effectiveness parameters.

For these experiments we use four labeled sets of flood-related tweets in German (DE), English (EN), Spanish (ES) and French (FR).
These were labeled by crowdsourcing workers via crowdsourcing as previously described; the question the annotators had to answer was ``is this message indicating that a flood is happening or about to happen?''
Hence, positive examples are the ones that are related to flood risk and impact, and negative examples are the ones that are not.

For each language and word-embedding source we performed three experiments: mono-lingual, cold-start, and warm-start.
We used two thirds of the labeled data for training the classifiers and the remaining third for testing. We keep the testing portion fixed across experiments.

In the \emph{mono-lingual} experiment we simply use labeled data in one language to predict the label for messages in the same language.
In the \emph{cold-start} experiment we train a classifier for a new language using only labeled data for other languages; for instance for automatically labeling tweets in Spanish, we use a classifier trained on labeled data for English, German, and French.
In the \emph{warm-start} experiment we use a set-up similar to the one of the cold-start experiment, but we add a limited number of messages (300) labeled in the target language.
Each experiment is done
once using GloVe embeddings
and
once using MUSE embeddings.
We report precision, recall, and F-measure for each experimental setup in Table~\ref{table-experiments}.

\begin{table}[htb]
\centering
\caption{Classification results for four languages (German, English, Spanish, and French). TL indicates the total number of labeled messages, while Pos. indicates the percentage of those who were labeled as flood-related. P, R, and F indicate Precision, Recall, and F-Measure respectively. We report the performance of a mono-lingual classifier, of a cross-language classifier with ``cold start'' (uses no training data in the target language), and of a cross-language classifier with ``warm start'' (uses 300 labeled items in the target language).}
\label{table-experiments}
\resizebox{\columnwidth}{!}{%
\begin{tabular}{lll|lllllllll|lllllllll}
\toprule
\multicolumn{3}{l}{\multirow{2}{*}{\begin{tabular}[c]{@{}l@{}}\end{tabular}}} & \multicolumn{9}{|c}{Glove embeddings} & \multicolumn{9}{|c}{MUSE embeddings} \\
\multicolumn{3}{l}{}
 & \multicolumn{3}{|c}{mono-lingual}
 & \multicolumn{3}{c}{cold-start}
 & \multicolumn{3}{c}{warm-start}
 & \multicolumn{3}{|c}{mono-lingual}
 & \multicolumn{3}{c}{cold-start}
 & \multicolumn{3}{c}{warm-start} \\
  & TL & Pos. & P & R & F & P & R & F & P & R & F & P & R & F & P & R & F & P & R & F \\
  \midrule
DE & 2356 & 46\% & 0.95 & 0.82 & 0.87  & 0.59 & 0.85 & 0.70 & 0.93 & 0.8 & 0.86 & 0.88 & 0.85 & 0.87 & 0.54 & 0.82 & 0.65 & 0.89 & 0.80 & 0.84 \\
EN & 1999 & 20\% & 0.79 & 0.63 & 0.70 & 0.59 & 0.49 & 0.54 & 0.67 & 0.51 & 0.58 & 0.64 & 0.68 & 0.66 & 0.33 & 0.50 & 0.40 & 0.58 & 0.28 & 0.38\\
ES & 1592 & 48\% & 0.80 & 0.78 & 0.79 & 0.61 & 0.75 & 0.67 & 0.71 & 0.83 & 0.77 & 0.70 & 0.84 & 0.77 & 0.62 & 0.69 & 0.65 & 0.68 & 0.89 & 0.77\\
FR & 1248 & 40\% & 0.74 & 0.72 & 0.73 & 0.50 & 0.46 & 0.48 & 0.62 & 0.77 & 0.69 & 0.69 & 0.75 & 0.72 & 0.44 & 0.86 & 0.58 & 0.59 & 0.72 & 0.65\\ \bottomrule
\end{tabular}
}%
\end{table}

Table~\ref{table-experiments} reports values for Precision (P) , Recall (R) and F-measure.
We also include the total number of tweets manually labeled by crowdsourcing workers (TL) and its percentage of positive (Pos.) tweets.
Collections for German, English, Spanish, and French consider tweets posted during floods happening in the last two years in Germany, the UK, Spain and Mexico, and France respectively.
Hence, the number of labeled tweets and the percentage of positive examples differ across languages.

All the results show the same pattern:
the mono-lingual classifier performs best, as expected;
the cold-start classifier (which does not use any labeled data in the target language) suffers from a loss mostly of precision, but also of recall;
and
the warm-start classifier (which involves annotating a small number of tweets in the target language) has better performance than the cold-start classifier both in terms of precision and recall.
Indeed, the warm-start classifier often achieves an F-measure that is comparable to the one of the mono-lingual classifier.
Regarding the choice of word embeddings, results suggest that the performance using GloVe or MUSE embeddings are comparable.

In general, considering the combination of the information from the classification with the known locations from the EFAS forecasts, the classification performance is sufficient to extract representative tweets from an event and to map approximately the affected locations, as we demonstrate next.

\subsection{Case Study: Calabria Floods in October 2018}
\label{subsec:experiments-demonstration}

In early October 2018, floods affected the region of Calabria in southern Italy. At least 2 people died in flash flooding after severe weather which peaked on October 5th.
A mother and her seven year old son, who were swept away by flood waters in their car, were found in a river near Lamezia Terme, between the towns of San Pietro a Maida and San Pietro Lametino in Calabria.
Other areas of Calabria were also hit by flooding and landslides.

\begin{figure}[htb!]
    \begin{minipage}{0.34\textwidth}\centering
        \includegraphics[width=\linewidth]{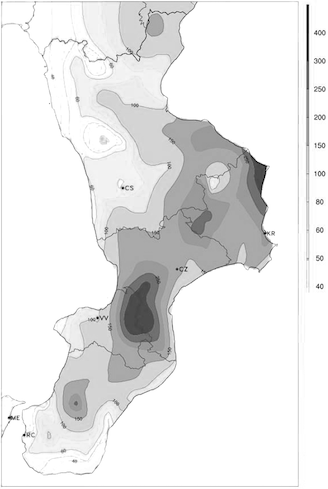}
        \textbf{(a) Rainfall}
    \end{minipage}
    \begin{minipage}{0.31\textwidth}\centering
        \includegraphics[width=\linewidth]{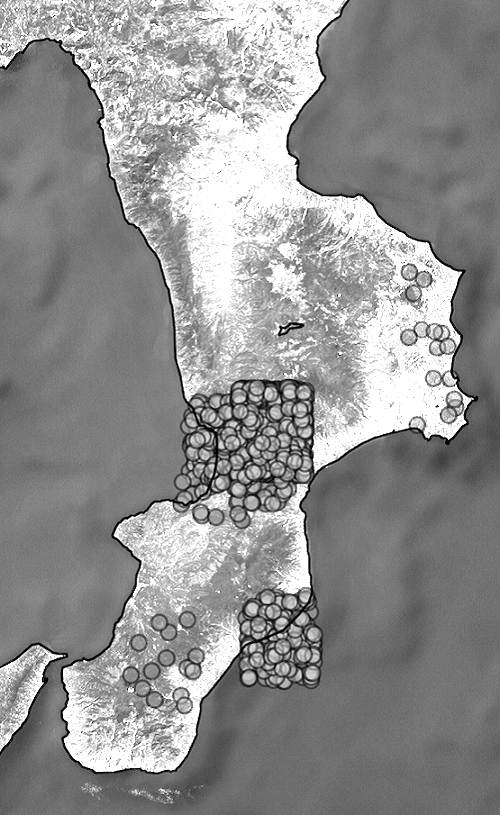}
        \textbf{(b) Tweets, cold start}
    \end{minipage}
    \begin{minipage}{0.31\textwidth}\centering
        \includegraphics[width=\linewidth]{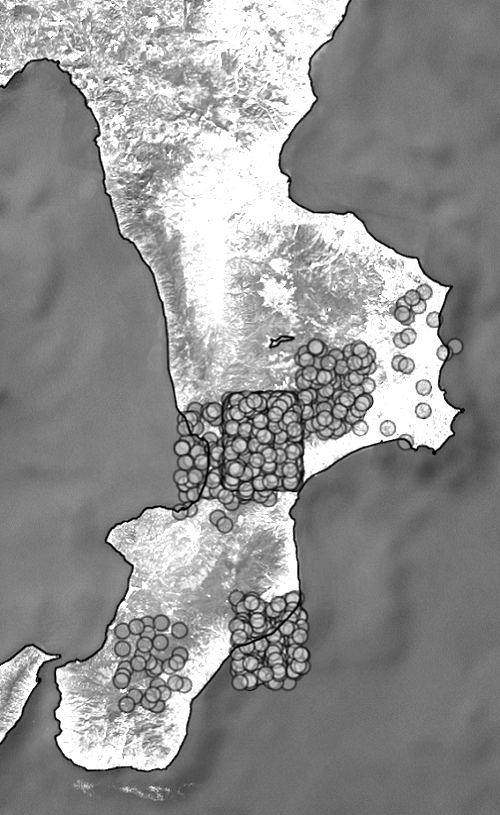}
        \textbf{(c) Tweets, warm start}
    \end{minipage}
\caption{Comparison of (a) rainfall, (b) tweets located by cold-start model, and (c) tweets located by warm-start model. Data from floods in Calabria, Italy, 2-5 October 2018. Tweets falling in the same location are randomly scattered for visualization purposes.}
\label{case-study-map}
\end{figure}

Several families were forced to evacuate their homes and people were rescued after they climbed onto the rooftops of houses to escape the flooding.\footnote{Data provided by \rurl{floodlist.com}, which is a EU-supported project providing reports on past floods.}
Italian news agency ANSA, stated that the Ponte delle Grazie bridge on provincial highway 19 in the area collapsed during the storms \parencite{ansa_2018}.
Vigili del Fuoco, Italy’s National Firefighters Corps, reported major flooding in Ciro Marina, Petilia de Policastro, Strongoli, Cotronei and Isola di Capo Rizzuto.
As shown in Figure~\ref{case-study-map}~(a) more than 300 mm of rain fell in 3 days \parencite{centrometeoitaliano_2018}.

EFAS forecasted a potential flood in the Calabria NUTS-2 area on the 4th of October with a predicted peak time of the event for the following day.
As planned, SMFR triggered a collection with a duration of 2 days that was later extended for an additional day due to persistence of the signal from EFAS forecasts. We analyzed the collection once it was stopped, at midnight on the 7th of October, after collecting 14,347 tweets.

In order to confirm what emerged from experiments in the previous section, we trained two classifiers for messages in Italian, the first (cold-start)
using only labeled data in German, English, Spanish, and French,
and the second (warm-start) adding 300 manually labeled tweets in Italian from the collected dataset.
For brevity we present results obtained using the GloVe embeddings (results using MUSE embeddings are similar).

In Figures \ref{case-study-map}~(b) and (c) we depict the position of geo-located tweets annotated by the cold-start and warm-start classifier respectively.
Tweets have been filtered using a relevance to flood (label predicted) greater or equal to $0.8$. We include the tweets geolocated within the bounding box used for triggering the collection, resulting in 2,847 tweets for the cold-start scenario against 3,857 for the warm-start scenario.
For visualization purposes, tweets geo-located to the exact same location are randomly scattered by a small amount in the map.

Figure~\ref{case-study-map} confirms the results from the experiments, in the sense that both cold-start and warm-start classifiers are able to classify relevant tweets, with an advantage for the warm-start classifier in the sense that it identifies more relevant tweets and has better coverage of the areas affected by heavy rainfall.
This suggests that the cold-start method can provide a first approximation for identifying an ongoing event, while the warm-start method yields more precise and relevant tweets.

Finally, Table~\ref{most-repr-tweets} shows samples of tweets that have been selected as the most representative for this event, following the heuristic described previously, and taking as input the tweets found relevant using the cold-start and warm-start methods.
We can see that in the cold-start scenario the most representative tweets are relevant to the event considered.
However, the warm-start scenario gives more informative messages.
While tweets selected by the cold-start classifier are relevant, they mostly just confirm the event as reported in the news; in constrast, the tweets selected in the warm-start case identify a message from the Italian Prime minister as the most representative (``I follow with concern the evolution of events ...'') and include information about damages and casualties due to the flood.
\begin{table}[htb!]
  \small
\caption{Representative tweets selected by cold-start and warm-start. Conf. is the confidence of the classifier. Mult. the multiplicity (number of near-duplicates of the tweet). Cent. is the centrality (number of closely related but not duplicate tweets).}
\label{most-repr-tweets}
\begin{tabular}{lllp{12cm}}
\toprule
\multicolumn{4}{c}{\textbf{Cold-start}} \\
\midrule
Conf. & Mult. & Cent. & Text (first $\sim$10 words)\\
\midrule
1.0 & 87 & 89 & Second flood in Calabria in 40 days. Devastation and 2 casualties ... \\
    &     &    & \emph{\small (Seconda inondazione in Calabria in soli 40 giorni. Devastazione e 2 vittime ...)} \\
1.0 & 11 & 93 & Bad weather in Calabria, the kennel is flooded ... \\
    &     &    &  \emph{\small (Maltempo in Calabria, il canile e~\textquotesingle sommerso dall\textquotesingle~acqua ...)} \\
1.0 & 7 & 94 & Bad weather: Red alert in Calabria today and in Puglia tomorrow ...\\
    &     &    &  \emph{\small (Maltempo: oggi allerta rossa in Calabria e domani in Puglia ...)} \\
1.0 & 5 & 97 & Meteo, panic in Calabria: streams flooding roads. Rescuers using rubber boats ...\\
    &     &    &  \emph{\small (Meteo, caos in Calabria: torrenti esondati e strade allagate. Soccorsi in gommone ...)} \\
1.0 & 5 & 87 & Bad weather in Calabria, missing mother and her two sons found dead ...\\
    &     &    &  \emph{\small (Maltempo Calabria, trovati morti mamma e due bimbi dispersi ...)} \\
\toprule
\multicolumn{4}{c}{\textbf{Warm-start}} \\
\midrule
Conf. & Mult. & Cent. & Text (first $\sim$10 words)\\
\midrule
1.0 & 194 & 76 & I follow with concern the evolution of events in \#Calabria ... \\
    &     &    & \emph{\small (Seguo con apprensione l~\textquotesingle~evolversi degli eventi in \#Calabria ...)} \\
1.0 & 194 & 88 & Water bomb in Calabria, among the upset in the population ...\\
    &     &    &  \emph{\small (Bomba d~\textquotesingle~acqua in Calabria, tra la popolazione sconvolta ...)} \\
1.0 & 14 & 46 & \#~breakingnews Bad weather Calabria: a woman and one of her son found dead. ...\\
    &     &    &  \emph{\small (\#ultimora Maltempo Calabria: morta una donna e suo figlio, disperso il fratello ...)} \\
1.0 & 23 & 98 & Bad weather in Calabria, mom and son found dead, missing 2yrs old brother ...\\
    &     &    &  \emph{\small (Maltempo in Calabria, morti mamma e figlio: sic erca il fratellino di 2 anni ...)} \\
1.0 & 8 & 94 & Bad weather, nigthmarish night in Calabria, Civil Protection: ``High risk'' ...\\
    &     &    &  \emph{\small (Maltempo, notte da incubo in Calabria, Protezione civile: ``rischio vittime''  ...)} \\
\bottomrule
\end{tabular}
\end{table}

\section{Conclusions and Future Work}
\label{sec:conclusions}

We have described how to integrate social media into the European Flood Awareness System (EFAS) to provide valuable signals augmenting flood risk information provided by this platform, by finding potentially relevant and representative messages from flood-affected areas in the languages spoken in those areas.
Our work provides a solution and methodology for integrating flood modeling and evidence from the ground in real-time for several countries, potentially providing information from local witnesses or local media to first responders. This unique combination of hydrological simulation forecasting and an automatic, immediate monitoring of the extent of the event through social media without necessity to manually translating information, allows to shorten the response time, which is extremely precious in the very early stages of a flood.
Moreover, during the development of an event, collected messages could be valuable to international rescue coordinators such as ERCC because they provide insights about the local response, about whether alerts that have been issued by authorities, and about some of the concerns that those affected by a flood or a flood alert may have.
The research also highlighted the need for high-recall data collection in which data in multiple languages is captured, and provided a methodology for dealing with a new language, by bootstrapping a classifier with similar languages for which labeled data is available, using either language-agnostic or language-aligned word embeddings.
Additionally, it was clear during the development of the project that naturally occurring data (i.e., actual messages posted during a flood in a particular country) are necessary to build an accurate classifier and aggregator.

\smallskip\noindent\textbf{Limitations.}
Due to the nature of EFAS and its geographical domain (pan-European) we focused on language spoken in the region, therefore we only tested the methodology proposed for indoeuropean languages. The performance with other languages remains to be studied.

At the time of writing, the described system is still in testing phase. After a period of internal evaluation, SMFR will become operational and its results will be disseminated among EFAS Partners.

\subsection{Future Work}

As future research activities, one can envision a global system comprising dozens of languages used to augment the Global Flood Awareness System (GloFAS; http://www.globalfloods.eu/). One can also envision further steps in the direction of using social media as a data source that can feed into a predictive model, using it not only for confirming known flood risks, but also for detecting new ones, particularly in areas where digital devices are prevalent but meteorological stations and other physical sensors are scarce in comparison.

\smallskip\noindent\textbf{Reproducibility.}
Datasets and code for the experiments described on this paper will be available for research purposes at \url{https://bitbucket.org/lorinivalerio/iscram\_2019/}.

\section{Acknowledgments}
This job has been funded by an Exploratory Research grant made available by the Joint Research Centre of the European Commission. Thanks to Gabriele Mantovani Banella and Sasa Vranic for helping in the development of web layers and web services for the integration of SMFR into the existing EFAS/GloFAS systems. Castillo thanks La Caixa project LCF/PR/PR16/11110009 for partial support.

\printbibliography


\end{document}